\documentstyle[11pt]{article}
\begin{document}

\title{On Stieltjes relations, Painlev\'e-IV hierarchy and complex monodromy}
\author{{\Large A.P. Veselov } \\
    Department of Mathematical Sciences, Loughborough University \\
    Loughborough, Leicestershire LE11 3TU, U.K.  \\
    and \\
    Landau Institute for Theoretical Physics, Moscow, Russia\\
    e-mail: {\tt A.P.Veselov@lboro.ac.uk} \\
  }
\date{}
\maketitle

\newtheorem{theorem}{Theorem}

\pagestyle{plain}

\maketitle

\begin{center}
 To the memory of J{\"u}rgen Moser
\end{center}

\begin{abstract}
A generalisation of the Stieltjes relations for the Painlev\'e-IV transcendents and
their higher analogues determined by the dressing chains is proposed.
It is proven that if a rational function from a certain class satisfies
these relations it must be a solution of some higher Painlev\'e-IV equation.
The approach is based on the interpretation of the Stieltjes relations
as local trivial monodromy conditions for certain Schr\"odinger equations
in the complex domain. As a corollary a new class of the Schr\"odinger operators
with trivial monodromy is constructed in terms of the Painlev\'e-IV transcendents.
\end{abstract}

\section{Introduction and formulation of the results.}
In 1885 Stieltjes \cite{S1,S2} has found the following remarkable interpretation of the zeroes of Hermite polynomials
$H_n(z)$:
$$H_n(z) = (-1)^n e^{z^2}\frac{d^n}{dz^n}e^{-z^2}.$$
Consider $n$ particles on the line interacting pairwise with repulsive logarithmic potential 
in the harmonic field. Then the equilibrium of this system is exactly the set of zeroes of $H_n(z)$.
More precisely, the extremum condition for the function
$$ U(z_1,...,z_n) = \sum_{j=1}^{n} z_j^{2} - \sum_{j<k}^{n} \ln(z_j - z_k)^2, $$
which is the system of the {\it Stieltjes relations}
\begin{equation}
\label{e}
\sum_{j \ne k}^{n} (z_k - z_j)^{-1} - z_k = 0, \\ k=1,...,n,
\end{equation}
determines exactly the roots of the equation $H_n(z) = 0.$
What is important for us is that although all these roots are real the result
is true in the complex domain as well, i.e. all the complex solutions of
the system (\ref{e}) are actually real and coincide with the zeroes of Hermite polynomial $H_n(z)$.
The proof is not difficult and based on the fact that $y = H_n(z)$ satisfies the 
following second-order linear differential equation
$$ y'' - 2z y' + 2n y = 0 $$ and this actually determines $H_n(z)$ up to a constant multiplier 
(see e.g. Szego's classical book \cite{Sz}, pp. 140-141). The sign $'$ here and below means
the derivative with respect to $z$.

The goal of this paper is to show that these relations have a natural analogue 
for the fourth Painlev\'e transcendents (and their higher analogues) and to explain how all this
is related to the theory of the Schr\"odinger operators with trivial monodromy in the complex domain.

To explain the relation of the Stieltjes result with the fourth Painlev\'e 
equation (PIV)
\begin{equation}
\label{PIV}
2w w'' = w'^2 + 3w^4 + 8zw^3 + 4(z^2-a)w^2 + 2b
\end{equation} 
let's first recall the well-known fact that the logarithmic derivative of the Hermite polynomial
$w = - (\ln H_n(z))^{'}$ satisfies PIV
with special parameters $a = (n+1), b = 2n^2$ (see \cite{L,G,BCH}).
Notice that the zeroes of Hermite polynomials are the simple poles
of the corresponding rational solution $w$ of PIV, each of them has
the residue -1.
The second remark is that the Stieltjes relations (\ref{e}) are equivalent to the fact
that the function $f = -(z + w) = - z + (\ln H_n(z))^{'}$ has no constant terms at the Laurent
expansions at all the poles or, equivalently,
that all the residues of the function
$f^2 = ((\ln H_n(z))' - z)^2$
are zeroes: 
$${\rm Res}\, f^{2}(z) = 0.$$
Indeed one can easily check that the left hand side of the relation (\ref{e}) 
is proportional to the corresponding constant term in the Laurent expansion of $f$
at the pole $z = z_k.$

Our first simple observation is that in this form this relation holds for any solution of 
PIV equation.

\begin{theorem}
For any solution $w$ of the fourth Painlev\'e equation
the residues of the function $(z + w)^2$ are
zero at all the poles of the solution $w$:
\begin{equation}
\label{Res}
{\rm Res}\, (z + w)^2 = 0.
\end{equation}

\end{theorem}
Actually we will prove a more general result about the following
system introduced by A.B.Shabat and the author in \cite{SV} in relation 
with the spectral theory of the Schr\"odinger 
operators under the name 
{\it dressing chain}:
\begin{equation}
\label{dc}
(f_i + f_{i+1})' = f_i^2  - f_{i+1}^2 + \alpha_i,\,\,i=1,2,...,N,
\end{equation}
where $N = 2p +1$ is odd, $\alpha_1, \alpha_2,..., \alpha_N$ are some constant parameters 
and we assume that $f_{N+1} = f_1$.
In \cite{SV} it was shown that this system has many remarkable
properties, in particular it passes the Kovalevskaya-Painlev\'e test. 

When $N=3$ and $\alpha = \sum_{i=1}^{N} \alpha_i = -2$
the dressing chain (\ref{dc}) is equivalent to PIV:
the function $w = -(z +f_1)$ satisfies the fourth Painlev\'e equation
(\ref{PIV}) with $a = \frac{1}{2}(\alpha_3 - \alpha_1), b = -\frac{1}{2}\alpha_2^2$
(see \cite{SV,A}). 

Similarly a dressing chain with any odd period $N \geq 3$ is equivalent to some 
nonlinear ordinary differential equation on $f = f_1$ of order $N$ (or $N-1$
if we fix the sum of $f_i$ to be $\alpha z$). Slightly abusing the terminilogy
we will call such equations the {\it higher PIV equations} which altogether form the {\it PIV hierarchy.}
We should mention that an equivalent hierarchy has been considered also by M. Noumi and Y. Yamada
\cite{NY1} who were not familiar with the theory of the dressing chain \cite{SV,A}.

\begin{theorem}
For any meromorphic solution of the dressing chain (\ref{dc}) the function $f=f_1$ 
has the poles of the first order with integer residues.
At any pole $z_0$ with ${\rm Res}_{z = z_0}\, f = m$ the following generalised 
Stieltjes relations are satisfied:
\begin{equation}
\label{f}
{\rm Res}_{z = z_0}\, f^2 = {\rm Res}_{z = z_0}\, f^4 = ... = {\rm Res}_{z = z_0}\, f^{2|m|} = 0.
\end{equation}
In particular, the residues of such function $f^2$ are
zero at all the poles of $f$:
\begin{equation}
\label{gRes}
{\rm Res}\, f^2 (z) = 0.
\end{equation}
\end{theorem}

The main question is how strong are these relations. We will show that
at least for the rational solutions they are indeed very strong and can be used 
as their characteristic property.

More precisely, let's consider a class of rational functions of the form
\begin{equation}
\label{int}
f = \sum_{i=1}^{n}\frac{m_i}{z-z_i} + \nu - \mu z,
\end{equation} 
where $m_i$ are some integers. 
It is easy to show (see \cite{SV} and section 2 below) that this is a general form of the
rational solutions of the dressing chains. In particular, for the PIV equation (\ref{PIV}) 
all the rational solutions have the form $w = -(f + z)$, 
where $f$ has a form (\ref{int}) with $m_i = \pm 1$, $\nu = 0$ and $\mu = \pm 1$ or $-1/3$ (see \cite{BCH}).

\begin{theorem}
If a rational function $f$ from the class (\ref{int}) satisfies the generalised Stieltjes relations
(\ref{f}) then $f$ is a rational solution of some higher Painleve-IV equation.
\end{theorem}

For a generic solution the residues $m_i = \pm 1$ (see section 2), so we have only the usual 
Stieltjes relations 
(\ref{gRes}), or explicitly \footnote
{As I have learnt from V.G.Marikhin in this form (with $m_i = \pm 1$ and $\nu = 0$) the Stieltjes relations
for the rational solutions of PIV equations 
have been written (under the name "Coulomb gas equations") in the recent paper \cite{MSBP}.
The question how strong these relations are was not addressed in this paper.}
\begin{equation}
\label{st}
\sum_{j \ne k}^{n}\frac{m_j}{z_k - z_j} + \nu - \mu z_k = 0, \,\, k=1,...,n
\end{equation} 
They can be represented as the extremum conditions for the function 
$$ V(z_1,...,z_n) =  \mu \sum_{j=1}^{n} m_j (z_j - \frac{\nu}{\mu})^{2} - \sum_{j<k}^{n} m_j m_k \ln(z_j - z_k)^2,$$
which also has a "physical" interpretation: it is the potential of the system of charged particles of charge $m_j$
with logarithmic pairwise interaction in an external harmonic field with charge $\mu$ centered at ${\nu}/{\mu}$. 
In case when all $m_i = 1, \nu = 0$ and $\mu = -1$ (or, equivalently, if all $m_i = -1, \nu = 0, \mu = 1$) 
we have the Stieltjes system. 

Actually we will describe all the solutions of the system (\ref{st}) explicitly as zeroes
of certain polynomials: Schur polynomials if $\mu = 0$ and the wronskians of Hermite polynomials if $\mu \ne 0$
(see Section 4 below).
The main idea is to interpret the Stieltjes relations as the trivial monodromy
conditions for certain Schr\"odinger operators in the complex domain and then to use
the known results about such operators \cite{DG}, \cite{O}.
More precisely, let us consider such an operator
$$ L = -\frac{d^2}{dz^2} + u(z)$$ 
with a potential $u$ which is meromorphic in the whole 
complex plane. We will say that the operator $L$ {\it has trivial monodromy}
if all the solutions of the corresponding Schr\"odinger equation
$$ L\psi = -\psi '' + u \psi = \lambda \psi $$
are also meromorphic in the whole complex plane for {\it all} $\lambda$.

Duistermaat and Gr\"unbaum were probably the first to consider the problem
of the classification of all such operators. In their fundamental paper on bispectrality \cite{DG}
they have solved this problem in the class of the rational potentials decaying at infinity.
Oblomkov \cite{O} recently generalised this result to the case when the potential
has a quadratic growth at infinity. Gesztesy and Weikard investigated 
the case of the potentials given by elliptic functions \cite {GW}.

It turns out that the Stieltjes relations (\ref{Res}) are exactly 
the local trivial monodromy conditions for the following new class 
of the Schr\"odinger operators related to PIV transcendents.
Let $w$ be any solution of PIV equation (\ref{PIV}) (which is known to be
meromorphic in the whole complex plane) and let's consider the Schr\"odinger operator $L$
with the potential
\begin{equation}
\label{u}
u = w' + (w + z)^2.
\end{equation}

\begin{theorem}
For any solution $w$ of the fourth Painleve equation (\ref{PIV}) the Schr\"odinger operator 
$L$ with the potential (\ref{u}) has trivial monodromy in the complex plane.
The same is true for the operators with the potentials 
$$u = f' + f^2,$$
where $f=f_1$ for any meromorphic solution of some dressing chain.
\end{theorem}

As a by-product we have the proof of following result first established by F.Calogero \cite{C} 
(see also \cite{P}).
The following system of {\it Calogero relations}
\begin{equation}
\label{C}
2 \sum_{j \ne k}^{n} (z_k - z_j)^{-3} - z_k = 0, \\ k=1,...,n,
\end{equation}
describes the equilibriums in the well-known Calogero-Moser model with the potential
$$ V_{CM}(z_1,...,z_n) = \sum_{j=1}^{n} z_j^{2} - 2 \sum_{j< k}^{n} (z_j - z_k)^{-2}. $$
{\bf Theorem 5.} (F.Calogero) 
{\it The Stieltjes relations (\ref{e}) imply the Calogero relations (\ref{C}).}

On the reals these relations are equivalent but in the complex domain Calogero relations 
have many more solutions different from the zeroes of Hermite polynomials (see sections 3 and 4 below).
For a discussion of the hierarchy of the similar relations for some classical
polynomials and Bessel functions we refer to a very interesting paper \cite{Ah}.

\section{Local analysis of the dressing chain and the proof of the Stieltjes relations
for PIV transcendents.}

Let's prove the theorem 2, the theorem 1 will then follow. 
The local expansions of the solutions of the dressing chain (\ref{dc}) near a pole 
(which we assume without loss of 
generality to be zero) have the form (see \cite{SV}):
$f_i = a_i z^{-1} + b_i + c_i z +... $. Substitution of this form into the 
system (\ref{dc}) gives an infinite system of equations for the coefficients.
The first two equations are
\begin{equation}
\label{1}
-(a_i + a_{i+1}) = a_i^2 - a_{i+1}^2
\end{equation}
and 
\begin{equation}
\label{2}
2a_ib_i - 2a_{i+1}b_{i+1} = 0.
\end{equation}
The first equation means that the coefficients $a_1, a_2,..., a_N$
determine a periodic trajectory of the 2-2 correspondence
\begin{equation}
\label{cor}
(x + y)(x - y + 1) = 0.
\end{equation}

{\bf Lemma 1.} {\it Any periodic trajectory of the correspondence (\ref{cor})
of an odd period must be integer and contain zero.}

Proof is elementary. Let's first show that such a trajectory must be integer.
This follows from the fact that the image of $x$ under the N-th iteration of the 
2-valued mapping (\ref{cor}) consists of the points of the form 
$x + k, -x + l$  with some integer $k,l$ such that $k \equiv N ({\rm mod}\, 2), l \equiv (N-1) ({\rm mod}\, 2).$
In particular, if $N$ is odd then $k \ne 0$ so for periodicity one can only
have $-x + l = x$, which means that $2x = l$, so $x$ is an integer since $l$
is an even number.

Now let's consider the absolute value of $x : u = |x|.$ Under the mapping (\ref{cor})
$u$ may either stay unchanged or move by 1 in either positive or negative direction.
Obviously for periodic trajectories we have an even number of the last movements.
This means that for an odd period we have an odd number of changes the sign
which only can be possible if at least one element of the trajectory vanishes.
This finishes the proof of the Lemma.

{\bf Remark.} Notice that actually we have proved more: it follows from the proof 
that any periodic trajectory of the period $N = 2p + 1$ consists of integers between $-p$ and $p$.
The extreme examples are $-p, -p+1, -p+2,...,-1, 0, 1,..., p-1, p$ and $1, -1, 0, 0, 0,..., 0.$
Only for the last sequence one has a family of solutions depending
on the maximal number of free parameters (see \cite{SV}).

Now from the Lemma it follows that at any pole of the solution of (\ref{dc}) at least
one of the coefficients $a_i$ must be zero. The relation
(\ref{2}) shows that the product $2a_ib_i = 2a_{i+1}b_{i+1}$ is independent of $i$
and because $a_i = 0$ for some $i$ this is zero for all $i$.
This means that if $a_j \ne 0$ than the corresponding $b_j=0$,
and therefore ${\rm Res}\, f_i^2 = 0$ for all $i$.
Continuing in a similar way one can prove that if $a_j = m$
then all the coefficients at the Laurent expansions of $f_j$
at even powers $z^{2k}$ are zeroes for all $k=0, 1,...,|m|-1.$
This implies theorem 2 (and therefore theorem 1).

The theorem 1 can be proven also directly from the local analysis of the 
PIV equation. Indeed, substituting the general form of the pole expansion of the
solution $w$ 
$$w = \alpha (z - z_0)^{-1} + \beta + \gamma (z - z_0) +...$$
into PIV equation (\ref{PIV}) one can easily derive that
\begin{equation}
\label{alpha}
\alpha = \pm 1
\end{equation}
and 
$$\beta = -z_0.$$
The last relation means that the constant term of the similar expansion
for the function $f = w + z$ at this pole is zero:
$$w + z = \alpha (z - z_0)^{-1} + (\beta + z_0) + (\gamma + 1)(z - z_0) +... =
\alpha (z - z_0)^{-1} + (\gamma + 1)(z - z_0) +...$$
This gives a direct proof of the Stieltjes relations for the general solution of the 
PIV equation. 

{\bf Remark.} For PIV equation there is a theorem saying that all the solutions are meromorphic
in the whole domain. We believe that the same is true for the solutions of the dressing chains 
(i.e. for the whole PIV hierarchy) but the proof is still to be found.

\section{Stieltjes and Calogero relations as trivial monodromy conditions.}

Now we are going to prove theorem 4 leaving the proof of the theorem 3 for the next section. 
Let us consider the Schr\"odinger equation
\begin{equation}
\label{ode}
-\varphi''+u(z)\varphi = \lambda \varphi
\end{equation}
with a meromorphic potential $u$ having poles only of second order.
Near such a pole (which can be assumed for simplicity to be $z = 0$)
the potential can be represented as Laurent series
$$u = \sum_{i = -2}^{\infty} c_{i} z^{i}.$$ 
Following the classical Frobenius approach (see e.g. \cite{Ince})
one can look for the solutions of the form 
$$ \varphi = z^{-\mu}(1 + \sum_{i=1}^{\infty} \xi_i z^{i}).$$
The corresponding $\mu$ must satisfy the characteristic equation
$\mu(\mu + 1) = c_{-2},$ which means that the equation (\ref{ode}) has a
meromorphic solution only if the coefficient $c_{-2}$ at any pole has
a very special form:
\begin{equation}
\label{loc1}
c_{-2} = m(m+1), m\in {\bf Z}_+.
\end{equation}
This condition is in fact not sufficient: the corresponding solution $\varphi$
may have a logarithmic term. A simple analysis shows (see e.g. \cite{DG})
that the logarithmic terms are absent for all $\lambda$ if and only if
in addition to (\ref{loc1}) all the first $m+1$ odd coefficients at the
Laurent expansion of the potential are vanishing:
\begin{equation}
\label{loc2}
c_{2k-1} = 0, k = 0, 1,..., m.
\end{equation}
 
The relation of this theory with the Stieltjes relations is explained by the following
simple but important Lemma.

{\bf Lemma 2.} {\it Let $f$ be a meromorphic function having the poles of the first order
with integer residues. The Schr\"odinger operator $L$ with the potential $u = f' + f^2$ 
has trivial monodromy in the complex domain if and only if at any pole $z_0$
with ${\rm Res}_{z = z_0}\, f = m$ the following relations are satisfied:}
$${\rm Res}_{z = z_0}\, f^2 = {\rm Res}_{z = z_0}\, f^4 = ... = {\rm Res}_{z = z_0}\, f^{2|m|} = 0.$$

Proof is straightforward: one can easily check by the substitution of 
$f = \frac{\pm m}{z-z_0} + \sum_{k=0}\alpha_k (z-z_0)^k$ with $m\in {\bf Z}_+$
into $u = f' + f^2$ that $c_{-2} = m(m \pm 1)$ and that
the trivial monodromy conditions $c_{2k-1} = 0, k = 0, 1,..., m-1$ 
are equivalent to the vanishing of the coefficients $\alpha_{2k} = 0, k = 0, 1,..., m-1.$
A remarkable fact is that an additional relation $c_{2m-1} = 0,$ which should be checked
for the negative residue $-m$ is then fulfilled automatically.

Combining this lemma with the first two theorems proved in the previous section
we come to theorem 4.

Now let us explain how this implies Calogero result.
Consider the rational function $w$ of the form
$$w = \sum_{i=1}^{n}\frac{-1}{z-z_i}.$$
As we have shown the Stieltjes relations imply the local 
trivial monodromy conditions for the potential
$u = w' + (w + z)^2.$ 
The function $u$ has a form
\begin{equation}
\label{um}
u = z^2 + \sum_{i=1}^{n}\frac{2}{(z-z_i)^2}
\end{equation}
since the residues $c_{-1}$ must be zero (see (\ref{loc2})).
It is easy to check that the second of the local trivial monodromy conditions $c_1 = 0$  
for such a potential are precisely the 
Calogero relations. This proves the theorem 5.

As we will see in the next section the Calogero relations holds not only
for the zeroes of the Hermite polynomials $H_k$ but also for the zeroes of all their wronskians
$W (H_{k_1}, H_{k_2},...,H_{k_n})$. This means that in the complex domain the Stieltjes relations 
actually are much stronger than the Calogero relations.

\section{Stieltjes relations and the rational solutions of the dressing chains.}

Consider now a rational function of the form (\ref{int}):
$$f = \sum_{i=1}^{n}\frac{m_i}{z-z_i} + \nu - \mu z,$$
where all $m_i$ are integers. We are going to describe all the functions of this form 
which satisfy the generalised Stieltjes relations (\ref{f}).
As a corollary we will prove the theorem 3 which claims that such a function must be a rational 
solution of a some higher PIV equation (\ref{dc}).

Notice first that the scaling transformations $f(z) \to \beta f(\beta z - \alpha)$ preserve
both the class of functions (\ref{int}) and the PIV hierarchy (\ref{dc}) 
Modulo these transformations we have
essentially only two different cases:  $\mu = 0$ (with an arbitrary $\nu$) and  $\mu = 1, \nu =0$.
Following the main idea of the previous section let's consider
the Schr\"odinger operator $L$ with the potential $u = f' + f^2.$
Lemma 2 says that the generalised Stieltjes relations (\ref{f})
implies that the operator $L$ has trivial monodromy in the whole complex plane.
Due to the relation ${\rm Res} f^2 = 0$ all the residues
of $u$ are zero so the potential has a form
$u = \sum_{i=1}^{n}\frac{2}{(z-z_i)^2} + c_0$ if $\mu = 0$ or
$u = z^2 + c_0 + \sum_{i=1}^{n}\frac{2}{(z-z_i)^2}$ if $\mu = 1$.

Now we can use the results of Duistermaat-Gr\"unbaum \cite{DG} and Oblomkov \cite{O}
which describe all such operators explicitly in terms of Darboux 
transformations.

Let's consider first the case $\mu = 0.$ Following Adler-Moser \cite{AM}
let's define the sequence of polynomials determined by the recurrence relation
$P_k(z)'' = P_{k-1}(z)$ with $P_1 = z$:
$$P_1 = z, P_2 = \frac{1}{6} z^3 + \tau_1, P_3 = \frac{1}{120} z^5 + \frac{1}{6} \tau_1 z^3 + \tau_2,...$$
and $W_n = W( P_1, P_2,..., P_n)$ be the Wronskian of these polynomials, which is also a polynomial in $z$ 
depending on $n$ additional parameters $\tau_1, \tau_2,..., \tau_n$.
This is a special case of Schur polynomials known also as Burchnall-Chaundy (or
Adler-Moser) polynomials. Duistermaat and Gr\"unbaum \cite{DG} have proved that if a 
rational potential $u$ decays at infinity and satisfies all the local trivial monodromy conditions
(\ref{loc1}),(\ref{loc2}) then (up to a shift $z \to z-a$) it must be  equal to the second logarithmic 
derivative of such polynomial $W_n$ with some parameters $\tau_1, \tau_2,..., \tau_n$:
$$u = - 2(\log W_n(z)).''$$
Corresponding $f$ is a rational solution of the Riccati equation $f' + f^2 = u$.
It is easy to show using the results of \cite{DG} that $f$ must be of the form
\begin{equation}
\label{0}
f =  \frac{d}{dz}\log \frac{W_{n \pm 1}}{W_n}
\end{equation}
if $\nu =0$ and 
\begin{equation}
\label{nu}
f =  \frac{d}{dz}\log \frac{\hat W_{n}}{W_n}, 
\end{equation}
if $\nu \ne 0.$ Here $\hat W_{n} =  W(P_1, P_2,..., P_n, e^{\nu z})$ is the Wronskian of the functions 
$P_1, P_2,..., P_n, e^{\nu z}$.

In the case $\mu = 1, \nu =0$ we have Oblomkov's generalisation of the Duistermaat - Gr\"unbaum result
which says that any Schr\"odinger operator with trivial monodromy and with the rational potential
growing at infinity as $z^2$ has the form
$$L = -\frac{d^2}{dz^2} - 2 \frac{d^2}{dz^2} \log W (H_{k_1}, H_{k_2},...,H_{k_n}) + z^2 + const,$$
where $H_k(z)$ is the $k$-th Hermite polynomial and $k_1, k_2, ..., k_n$ is a
sequence of different positive integers (see \cite{O}).
The corresponding functions $f$ have the form
\begin{equation}
\label{1}
f = \frac{d}{dz}\log \frac{ W (H_{k_1}, H_{k_2},...,H_{k_n}, H_{k_{n+1}})}{ W (H_{k_1}, H_{k_2},...,H_{k_n})} - z,
\end{equation}
where $k_1, k_2, ..., k_n, k_{n+1}$ are again some different positive integers.

The formulas (\ref{0}), (\ref{nu}), (\ref{1}) give a complete description (modulo natural affine transformations) 
of the functions $f$ of the form (\ref{int}) which satisfy the Stieltjes relations
(\ref{f}). Now we are ready to prove theorem 3. Indeed we have seen that for any such $f$ the corresponding
Schr\"odinger operator $L$ is a result of some number $m$ of the rational Darboux transformations applied 
to the operator $L_0 = -\frac{d^2}{dz^2} + \mu^2 z^2$.
Reversing this procedure we come back to $L_0$, then by taking $f_0 = -\mu z$ we can do one more step
which only shifts $L_0$ by a constant. Now we can apply our rational Darboux transformations to return
to the initial operator $L$. Thus we have constructed a closed chain of the rational Darboux transformations
of an odd length $N = 2m + 1,$ which is equivalent to the rational solution of the dressing chain (\ref{dc}) 
with $f_1 = f$ (see \cite{SV}). Theorem 3 is proven.

{\bf Remark.} To identify all $f$ of the form (\ref{1}) which satisfy some higher PIV of a {\it given} order
is actually a non-trivial task. In this relation I would like to mention the paper by M. Noumi 
and Y. Yamada \cite{NY2} where the rational solutions for the ordinary PIV equation ($N = 3$) 
have been described in terms the Schur functions for the special Young diagrams.

\section{Some open questions.}

We have seen that the rational solutions of the PIV hierarchy can be characterised as certain
rational functions satisfying the generalised Stieltjes relations (see Theorem 3).
It is natural to conjecture that these relations characterise also the general solutions of PIV hierarchy 
among all the meromorphic functions of certain order (in the sense of Nevanlinna) with integer residues. 
Rod Halburd suggested recently some interesting ideas which may help to prove this.
As an intermediate case one can consider
the special solutions of PIV equations expressed in terms of Weber-Hermite 
functions (see e.g. \cite{BCH}). 

Another interesting question: what are the analogues of the Stieltjes relations for other Painlev\'e
transcendents ? For example, for the second Painlev\'e equation PII
$$y'' = 2 y^3 + z y + a$$
one can easily see that the constant terms in the Laurent expansions
at the poles of the solutions must be zero, so we have
the relations 
$${\rm Res}\, y^2 = 0,$$ 
which can be considered as such an analogue. 

One more intriguing problem is to understand what is a proper multidimensional analogue
of the Stieltjes relations in the theory of multidimensional Baker-Akhiezer functions 
recently developed in \cite{CFV}.

\section{Acknowledgments.}
I came to the idea of this paper in 1992 when we have been working with A.B. Shabat
and V.E.Adler on the theory of the dressing chain.
I am very grateful to Rod Halburd for numerous stimulating discussions which convinced
me that such a paper should be written. I am also grateful to Alberto Gr\"unbaum 
for inspiring discussions of the beautiful Stieltjes results and to John Gibbons
who helped me to find the original Calogero paper \cite{C} containing an important extension of 
these results.

This research has been partially supported by EPSRC (grant No GR/M69548).

\end{document}